\documentclass[notitlepage,10pt,aps,amsmath,amssymb,eqsecnum,nofootinbib]{revtex4-1}

\usepackage{array}
\usepackage{bbm}

\usepackage{graphicx}

\usepackage{braket}

\newcommand{\be}{\begin{equation}}
\newcommand{\ee}{\end{equation}}
\newcommand{\bea}{\begin{eqnarray}}
\newcommand{\eea}{\end{eqnarray}}
\newcommand{\bi}{\begin{itemize}}
\newcommand{\ei}{\end{itemize}}

\newcommand{\bspl}{\begin{split}}
\newcommand{\espl}{\end{split}}

\def\mev{{\rm MeV}}
\def\gev{{\rm GeV}}
\def\tev{{\rm TeV}}

\def\ev{\mathrm{e\kern-0.1em V}}
\def\kev{\mathrm{ke\kern-0.1em V}}
\def\mev{\mathrm{Me\kern-0.1em V}}
\def\gev{\mathrm{Ge\kern-0.1em V}}
\def\tev{\mathrm{Te\kern-0.1em V}}

\newcommand{\RomatreINFN}{Istituto Nazionale di Fisica Nucleare, Sezione di Roma Tre,\\ Via della Vasca Navale 84, I-00146 Rome, Italy}
\newcommand{\Romatre}{Dipartimento di Matematica e Fisica and INFN, Universit\`a Roma Tre,\\ Via della Vasca Navale 84, I-00146 Rome, Italy}

\begin{document}

\title{Extraction of $\lvert V_{cd} \rvert$ and $\lvert V_{cs} \rvert$ from experimental decay rates\\ using lattice QCD \boldmath$D \to \pi(K) \ell \nu$ form factors}

\author{L.~Riggio} \affiliation{\RomatreINFN} 
\author{G.~Salerno} \affiliation{\Romatre}
\author{S.~Simula} \affiliation{\RomatreINFN} 

\begin{abstract}
We present a determination of the Cabibbo-Kobayashi-Maskawa matrix elements $|V_{cd}|$ and $|V_{cs}|$ obtained by combining the momentum dependence of the semileptonic vector form factors $f_+^{D \to \pi}(q^2)$ and $f_+^{D \to K}(q^2)$, recently determined from lattice QCD simulations, with the differential rates measured for the semileptonic $D \to \pi \ell \nu$ and $D \to K \ell \nu$ decays. 
Our analysis is based on the results for the semileptonic form factors produced by the European Twisted Mass Collaboration with $N_f = 2 + 1 + 1$ flavors of dynamical quarks in the whole range of values of the squared 4-momentum transfer accessible in the experiments. 
The statistical and systematic correlations between the lattice data as well as those present in the experimental data are properly taken into account.
With respect to the standard procedure based on the use of only the vector form factor at zero 4-momentum transfer, we obtain more precise and consistent results: $|V_{cd} |= 0.2341 ~ (74)$ and $|V_{cs} |= 0.970 ~ (33)$.
The second-row CKM unitarity is fulfilled within the current uncertainties:  $|V_{cd}|^2 + |V_{cs}|^2 + |V_{cb}|^2 = 0.996 ~ (64)$.
Moreover, using for the first time hadronic inputs determined from first principles, we have calculated the ratio of the semileptonic $D \to \pi(K)$ decay rates into muons and electrons, which represent a test of lepton universality within the SM, obtaining in the isospin-symmetric limit of QCD: ${\cal{R}}_{LU}^{D\pi} = 0.985~(2)$ and ${\cal{R}}_{LU}^{DK} = 0.975~(1)$.
\end{abstract}

\maketitle

\section{Introduction}
\label{sec:intro}

Weak decays of hadrons represent an important source of direct information about the Cabibbo-Kobayashi-Maskawa (CKM)~\cite{Cabibbo:1963yz,Kobayashi:1973fv} matrix elements, which are fundamental parameters describing the quark flavor mixing in the electroweak sector of the Standard Model (SM).
The CKM matrix is unitary in the SM and this gives rise to unitarity conditions between the elements of its rows and columns, whose investigation has been the focus of much of the experimental and theoretical efforts in Flavor Physics during the recent years and may provide stringent consistency tests of the SM. 

The golden modes for the determination of the CKM entries $|V_{cd}|$ and $|V_{cs}|$ are represented by leptonic and semileptonic decays of $D$ and $D_s$ mesons, which are sensitive probes of the $c \to d$ and $c \to s$ quark transitions.
According to the $V - A$ structure of the SM weak currents, leptonic and semileptonic $D$ and $D_s$ decays should provide consistent results for the CKM elements $|V_{cd}|$ and $|V_{cs}|$.

As is well known, the extraction of the CKM matrix elements from the experimental decay rates requires theoretical inputs, namely the leptonic decay constants and the semileptonic form factors, which encode the non-perturbative QCD dynamics and can be determined from first principles by simulating QCD on a lattice.
The current Lattice QCD (LQCD) determinations of the leptonic decay constants $f_D$ and $f_{D_s}$ and of the semileptonic vector form factors at zero four-momentum transfer $f_+^{D \to \pi}(0)$ and $f_+^{D \to K}(0)$ are summarized in the latest FLAG review \cite{Aoki:2016frl}.
Uncertainties equal to $2.4 \%$ and $1.7 \%$ are quoted respectively for $|V_{cd}|$ and $|V_{cs}|$ extracted from the leptonic decays, while larger uncertainties equal to $4.5 \%$ and $2.6 \%$, presently dominated by the theoretical errors, affect the determinations of $|V_{cd}|$ and $|V_{cs}|$ from semileptonic decays.

Within the SM the semileptonic $D \to P \ell \nu$ differential decay rate is given by
 \be
      \frac{d\Gamma(D \to P \ell \nu)}{d q^2} = \frac{G_F^2 |V_{cx}|^2}{24 \pi^3} \frac{(q^2 - m_\ell^2)^2 |\vec{p}_P|}{q^4 M_D^2} 
                                                                        \bigg[ \left( 1 + \frac{m_\ell^2}{2q^2} \right) M_D^2 |\vec{p}_P|^2 |f_+^{D P}(q^2)|^2 +
                                                                        \frac{3m_\ell^2}{8q^2}(M_{D}^2-M_P^2)^2|f_0^{D P}(q^2)|^2 \bigg] , ~
      \label{eq:Gamma}
 \ee
where $x = d (s)$ is the daughter quark, $|\vec{p}_P|$ is the momentum of the daughter $P = \pi (K)$ pseudoscalar meson in the $D$-meson rest frame and $q = ( p_D - p_P )$ is the 4-momentum of the outgoing lepton pair.
Since in Eq.~(\ref{eq:Gamma}) the contribution of the scalar form factor $f_0(q^2)$ is proportional to $m_\ell^2$, in the case of $\ell = e (\mu)$ final leptons the differential decay rate simplifies to 
 \be
    \label{eq:SimplifiedGamma}
    \frac{d\Gamma(D \to P e(\mu) \nu)}{d q^2} = \frac{G_F^2 \left| V_{cx} \right|^2}{24 \pi ^3} \left| \vec{p}_P \right|^3 \left| f_+^{D P}(q^2) \right|^2 . 
 \ee

The analyses of the experimental data from BELLE \cite{Widhalm:2006wz}, BABAR \cite{Lees:2014ihu,Aubert:2007wg}, CLEO \cite{Besson:2009uv} and BESIII \cite{Ablikim:2015ixa,Ablikim:2017lks} on $D \to \pi(K) e(\mu) \nu$ decays are based on a variety of parameterizations for the shape of the vector form factor $f_+^{D \pi(K)}(q^2)$, like effective pole models, inspired by either dispersion relations \cite{Becher:2005bg} or heavy-quark expansion arguments \cite{Becirevic:1999kt}, the $z$-expansion method \cite{Boyd:1997qw} or even relativistic quark model predictions \cite{Scora:1995ty}.
Thus, each experiment provides results for the products $|V_{cx}| f_+^{D P}(q^2)$ in various $q^2$-bins.

The LQCD calculations of the semileptonic $D \to \pi(K)$ form factors, fulfilling the quality criteria of the latest FLAG summary~\cite{Aoki:2016frl}, provide the value of the vector form factor at zero 4-momentum transfer, $f_+^{D\pi(K)}(q^2 = 0)$. 
This can be used to extract the relevant CKM entry adopting the experimental (averaged) values for the products $|V_{cd} |f_+^{D \to \pi}(0)$ and $|V_{cs} |f_+^{D \to K}(0)$, like the ones determined by the Heavy Flavor Averaging Group (HFAG) in Ref.~\cite{Amhis:2016xyh} by combining all the experiments.

Recently \cite{Lubicz:2017syv} the momentum dependence of both the vector and scalar form factors $f_+^{D\pi(K)}(q^2)$ and $f_0^{D\pi(K)}(q^2)$ at the physical pion mass and in the continuum and infinite volume limits has been calculated on the lattice by the European Twisted Mass Collaboration (ETMC), obtaining an overall agreement with the momentum dependence of the experimental data from BELLE \cite{Widhalm:2006wz}, BABAR \cite{Lees:2014ihu,Aubert:2007wg}, CLEO \cite{Besson:2009uv} and BESIII \cite{Ablikim:2015ixa}.
Some deviations have been nevertheless observed at high values of $q^2$.
At variance with the LQCD calculations considered in the FLAG review~\cite{Aoki:2016frl} the ETMC calculations of Ref.~\cite{Lubicz:2017syv} employ gauge configurations with $N_f = 2 + 1 + 1$ dynamical quarks (which include in the sea, besides two light mass-degenerate quarks, also the strange and charm quarks with masses close to their physical values), and provide also a set of synthetic (correlated) data points for the vector and scalar form factors in the whole range of values of $q^2$ accessible in the experiments, i.e.~from $q^2 = 0$ up to $q^2_{\rm max} = (M_D - M_{\pi(K)})^2$.

It is the aim of this work to combine the new LQCD results of Ref.~\cite{Lubicz:2017syv} for the vector form factor $f_+^{D\pi(K)}(q^2)$ with the experimental data on the differential rates of the semileptonic $D \to \pi(K) e(\mu) \nu$ decays in order to perform a consistent analysis within the SM for the extraction of the CKM matrix elements $|V_{cd}|$ and $|V_{cs}|$\footnote{A similar analysis aimed at the extraction of $|V_{cs}|$ from the semileptonic $D \to K \ell \nu$ data was attempted in the unpublished work~\cite{Koponen:2013tua} (see, however, Refs.~\cite{Aoki:2013ldr,Aoki:2016frl}).}.
We show that a more precise and consistent determination of $|V_{cd}|$ and $|V_{cs}|$, compared to the one based on the use of only the theoretical form factor values at $q^2 = 0$, can be obtained.

\section{Extraction of $|V_{cd}|$ and $|V_{cs}|$}
\label{sec:extraction}

The starting point is the partial decay rate provided by each experiment for various bins of values of $q^2$ (i.e., $q_i^2 \pm  \Delta q_i^2 / 2$ for $i = 1, ~ ..., ~ N_{bins}$).
By integrating Eq.~(\ref{eq:SimplifiedGamma}) in each experimental bin one has
\be
    \label{eq:integratedgamma}
    \left[ \Delta \Gamma(q_i^2) \right]^{\rm EXP} \equiv \int_{\Delta q_i^2} dq^2 \frac{d\Gamma(D \to P \ell \nu)}{dq^2} =
                                                                                       \frac{G_F^2 \left| V_{cx} \right|^2}{24 \pi^3} I(q_i^2) , 
\ee
where the r.h.s.~contains the phase-space integral over the vector form factor, viz.
 \be
    I(q_i^2) \equiv \int_{\Delta q_i^2} d q^2 \left| \vec{p}_P \right|^3 \left| f_+^{D P}(q^2) \right|^2 .
    \label{eq:lattintegral}
 \ee

Using the results of Ref.~\cite{Lubicz:2017syv} for the vector form factor $f_+^{D P}(q^2)$, we can combine the theoretical prediction $\left[ I(q_i^2) \right]^{\rm LAT}$ of Eq.~(\ref{eq:lattintegral}) with the experimental measurement of the partial decay rate (\ref{eq:integratedgamma}) and get a determination of $|V_{cx}|$ for each experimental $q^2$-bin:
\be
    \left| V_{cx}(q_i^2) \right|^2 = \frac{24 \pi^3}{G_F^2} \frac{\left[ \Delta \Gamma (q_i^2) \right]^{\rm EXP}}{\left[ I(q_i^2) \right]^{\rm LAT}} .
    \label{eq:Vcxq2bin}
\ee

The values of $\left[ \Delta \Gamma (q_i^2) \right]^{\rm EXP}$ measured for the $D \to \pi$ and $D \to K$ semileptonic decays by BABAR~\cite{Lees:2014ihu,Aubert:2007wg}, CLEO~\cite{Besson:2009uv} and BESIII~\cite{Ablikim:2015ixa} collaborations are collected in Tables~\ref{tab:exp_data_gammaDpi_BABAR}-\ref{tab:exp_data_gammaDK_BABAR}, ~\ref{tab:exp_data_gammaDpi_CLEO}-\ref{tab:exp_data_gammaDK_CLEO} and~\ref{tab:exp_data_gammaDpi_BESIII}-\ref{tab:exp_data_gammaDK_BESIII}, respectively.

\begin{table}[htb!]
\parbox{0.45\linewidth}{
\centering
\begin{tabular}{|ccc|}
\hline
Experiment & $q^2 ~ (\mbox{GeV}^2)$ & $\left[ \Delta \Gamma(q^2) \right]^{\rm EXP} \cdot 10^{16} ~ (\mbox{GeV})$ \\ \hline \hline
 BABAR& $(0.00,\,0.30)$ & ~ $8.09 ~\pm~ 0.46$ ~ \\ 
            & $(0.30,\,0.60)$ & ~ $7.51 ~\pm~ 0.59$ ~ \\ 
            & $(0.60,\,0.90)$ & ~ $7.31 ~\pm~ 0.53$ ~ \\ 
            & $(0.90,\,1.20)$ & ~ $6.15 ~\pm~ 0.46$ ~ \\ 
            & $(1.20,\,1.50)$ & ~ $4.88 ~\pm~ 0.46$ ~ \\ 
            & $(1.50,\,1.80)$ & ~ $4.28 ~\pm~ 0.46$ ~ \\ 
            & $(1.80,\,2.10)$ & ~ $3.39 ~\pm~ 0.46$ ~ \\ 
            & $(2.10,\,2.40)$ & ~ $1.98 ~\pm~ 0.40$ ~ \\ 
            & $(2.40,\,2.70)$ & ~ $0.77 ~\pm~ 0.33$ ~ \\ 
            & $(2.70,\,2.98)$ & ~ $0.14 ~\pm~ 0.13$ ~ \\ \hline 
 \end{tabular}
\vspace{0.15cm}
\caption{\it Values of the partial decay rates $\left[ \Delta \Gamma (q^2) \right]^{\rm EXP}$ for the $D \to \pi$ transition in the $q^2$-bins measured by BABAR~\cite{Lees:2014ihu}.}
\label{tab:exp_data_gammaDpi_BABAR}
}
\quad
\parbox{0.45\linewidth}{
\centering
\begin{tabular}{|ccc|}
\hline
Experiment & $q^2 ~ (\mbox{GeV}^2)$ & $\left[ \Delta \Gamma(q^2) \right]^{\rm EXP} \cdot 10^{15} ~ (\mbox{GeV})$ \\ \hline \hline
 BABAR& $(0.00,\,0.20)$  & ~ $11.69  ~\pm~ 0.34 $ ~ \\ 
             & $(0.20,\,0.40)$  & ~ $10.72  ~\pm~ 0.32 $ ~ \\ 
             & $(0.40,\,0.60)$  & ~ $ 9.50  ~\pm~ 0.27 $ ~ \\ 
             & $(0.60,\,0.80)$  & ~ $ 8.16  ~\pm~ 0.24 $ ~ \\ 
             & $(0.80,\,1.00)$  & ~ $ 6.53  ~\pm~ 0.20 $ ~ \\ 
             & $(1.00,\,1.20)$  & ~ $ 5.09  ~\pm~ 0.16 $ ~ \\ 
             & $(1.20,\,1.40)$  & ~ $ 3.51  ~\pm~ 0.13 $ ~ \\ 
             & $(1.40,\,1.60)$  & ~ $ 2.14  ~\pm~ 0.08 $ ~ \\ 
             & $(1.60,\,1.80)$  & ~ $ 0.85  ~\pm~ 0.06 $ ~ \\ 
             & $(1.80,\,1.88)$  & ~ $ 0.041 ~\pm~ 0.004$ ~ \\ \hline       
\end{tabular}
\vspace{0.15cm}
\caption{\it Values of the partial decay rates $\left[ \Delta \Gamma (q^2) \right]^{\rm EXP}$ for the $D \to K$ transition in the $q^2$-bins measured by BABAR~\cite{Aubert:2007wg}.}
\label{tab:exp_data_gammaDK_BABAR}
}
\end{table}

\begin{table}[htb!]
\parbox{0.45\linewidth}{
\centering
\begin{tabular}{|ccc|}
\hline
Experiment & $q^2 ~ (\mbox{GeV}^2)$ & $\left[ \Delta \Gamma(q^2) \right]^{\rm EXP} \cdot 10^{16} ~ (\mbox{GeV})$ \\ \hline \hline
 CLEO $D^0$ & $(0.00,\,0.30)$ & ~ $9.16 ~\pm~ 0.66$ ~ \\ 
            & $(0.30,\,0.60)$ & ~ $8.04 ~\pm~ 0.59$ ~ \\ 
            & $(0.60,\,0.90)$ & ~ $6.72 ~\pm~ 0.53$ ~ \\ 
            & $(0.90,\,1.20)$ & ~ $6.46 ~\pm~ 0.53$ ~ \\ 
            & $(1.20,\,1.50)$ & ~ $5.21 ~\pm~ 0.46$ ~ \\ 
            & $(1.50,\,2.00)$ & ~ $5.54 ~\pm~ 0.46$ ~ \\ 
            & $(2.00,\,2.98)$ & ~ $5.27 ~\pm~ 0.46$ ~ \\ 
            && ~ \\
            && ~ \\ \hline 
 CLEO $D^+$ & $(0.00,\,0.30)$ & ~ $4.68 ~\pm~ 0.46$ ~ \\ 
            & $(0.30,\,0.60)$ & ~ $4.35 ~\pm~ 0.46$ ~  \\ 
            & $(0.60,\,0.90)$ & ~ $3.69 ~\pm~ 0.46$ ~  \\ 
            & $(0.90,\,1.20)$ & ~ $3.76 ~\pm~ 0.46$ ~  \\ 
            & $(1.20,\,1.50)$ & ~ $3.16 ~\pm~ 0.46$ ~  \\ 
            & $(1.50,\,2.00)$ & ~ $3.56 ~\pm~ 0.46$ ~  \\ 
            & $(2.00,\,3.01)$ & ~ $2.44 ~\pm~ 0.46$ ~  \\ 
            && ~ \\
            && ~ \\ \hline 
\end{tabular}
\vspace{0.15cm}
\caption{\it Values of the partial decay rates $\left[ \Delta \Gamma (q^2) \right]^{\rm EXP}$ for the $D \to \pi$ transition in the $q^2$-bins measured by CLEO~\cite{Besson:2009uv}, given separately for the $D^0 \to \pi^- \ell \nu$ and $D^+ \to \pi^0 \ell \nu$ channels.}
\label{tab:exp_data_gammaDpi_CLEO}
}
\quad
\parbox{0.45\linewidth}{
\centering
\begin{tabular}{|ccc|}
\hline
Experiment & $q^2 ~ (\mbox{GeV}^2)$ & $\left[ \Delta \Gamma(q^2) \right]^{\rm EXP} \cdot 10^{15} ~ (\mbox{GeV})$ \\ \hline \hline
 CLEO $D^0$  & $(0.00,\,0.20)$  & ~ $11.74  ~\pm~ 0.28 $ ~ \\ 
             & $(0.20,\,0.40)$  & ~ $10.43  ~\pm~ 0.26 $ ~ \\ 
             & $(0.40,\,0.60)$  & ~ $ 9.17  ~\pm~ 0.23 $ ~ \\ 
             & $(0.60,\,0.80)$  & ~ $ 7.70  ~\pm~ 0.21 $ ~ \\ 
             & $(0.80,\,1.00)$  & ~ $ 6.17  ~\pm~ 0.19 $ ~ \\ 
             & $(1.00,\,1.20)$  & ~ $ 4.67  ~\pm~ 0.16 $ ~ \\ 
             & $(1.20,\,1.40)$  & ~ $ 3.52  ~\pm~ 0.13 $ ~ \\ 
             & $(1.40,\,1.60)$  & ~ $ 2.04  ~\pm~ 0.10 $ ~ \\              
             & $(1.60,\,1.88)$  & ~ $ 0.84  ~\pm~ 0.08 $ ~ \\ \hline 
 CLEO $D^+$  & $(0.00,\,0.20)$  & ~ $11.72  ~\pm~ 0.43 $ ~ \\ 
             & $(0.20,\,0.40)$  & ~ $10.29  ~\pm~ 0.39 $ ~ \\ 
             & $(0.40,\,0.60)$  & ~ $ 9.24  ~\pm~ 0.36 $ ~ \\ 
             & $(0.60,\,0.80)$  & ~ $ 8.09  ~\pm~ 0.32 $ ~ \\ 
             & $(0.80,\,1.00)$  & ~ $ 5.88  ~\pm~ 0.27 $ ~ \\ 
             & $(1.00,\,1.20)$  & ~ $ 5.38  ~\pm~ 0.24 $ ~ \\ 
             & $(1.20,\,1.40)$  & ~ $ 3.27  ~\pm~ 0.18 $ ~ \\ 
             & $(1.40,\,1.60)$  & ~ $ 1.76  ~\pm~ 0.12 $ ~ \\              
             & $(1.60,\,1.88)$  & ~ $ 0.78  ~\pm~ 0.09 $ ~ \\ \hline           
\end{tabular}
\vspace{0.15cm}
\caption{\it Values of the partial decay rates $\left[ \Delta \Gamma (q^2) \right]^{\rm EXP}$ for the $D \to K$ transition in the $q^2$-bins measured by  CLEO~\cite{Besson:2009uv}, given separately for the $D^0 \to K^- \ell \nu$ and $D^+ \to K^0 \ell \nu$ channels.}
\label{tab:exp_data_gammaDK_CLEO}
}
\end{table}

\begin{table}[htb!]
\parbox{0.475\linewidth}{
\centering
\begin{tabular}{|ccc|}
\hline
Experiment & $q^2 ~ (\mbox{GeV}^2)$ & $\left[ \Delta \Gamma(q^2) \right]^{\rm EXP} \cdot 10^{16} ~ (\mbox{GeV})$ \\ \hline \hline
 BESIII $D^0$& $(0.00,\,0.20)$ & ~ $6.14 ~\pm~ 0.27$ ~   \\ 
            & $(0.20,\,0.40)$ & ~ $5.38 ~\pm~ 0.26$ ~  \\ 
            & $(0.40,\,0.60)$ & ~ $4.81 ~\pm~ 0.25$ ~  \\ 
            & $(0.60,\,0.80)$ & ~ $4.89 ~\pm~ 0.24$ ~  \\ 
            & $(0.80,\,1.00)$ & ~ $4.83 ~\pm~ 0.25$ ~  \\ 
            & $(1.00,\,1.20)$ & ~ $4.28 ~\pm~ 0.23$ ~  \\ 
            & $(1.20,\,1.40)$ & ~ $3.74 ~\pm~ 0.22$ ~  \\ 
            & $(1.40,\,1.60)$ & ~ $3.17 ~\pm~ 0.20$ ~  \\ 
            & $(1.60,\,1.80)$ & ~ $3.08 ~\pm~ 0.20$ ~  \\ 
            & $(1.80,\,2.00)$ & ~ $2.42 ~\pm~ 0.17$ ~  \\ 
            & $(2.00,\,2.20)$ & ~ $1.86 ~\pm~ 0.15$ ~  \\ 
            & $(2.20,\,2.40)$ & ~ $1.32 ~\pm~ 0.13$ ~  \\ 
            & $(2.40,\,2.60)$ & ~ $0.75 ~\pm~ 0.10$ ~  \\ 
            & $(2.60,\,2.98)$ & ~ $0.62 ~\pm~ 0.09$ ~  \\ 
            && ~ \\
            && ~ \\
            && ~ \\
            && ~ \\ \hline 
 BESIII $D^+$ & $(0.00,\,0.30)$ & ~ $4.38 ~\pm~ 0.21$ ~ \\ 
            & $(0.30,\,0.60)$ & ~ $3.81 ~\pm~ 0.22$ ~  \\ 
            & $(0.60,\,0.90)$ & ~ $3.12 ~\pm~ 0.21$ ~  \\ 
            & $(0.90,\,1.20)$ & ~ $3.14 ~\pm~ 0.21$ ~  \\ 
            & $(1.20,\,1.50)$ & ~ $2.85 ~\pm~ 0.21$ ~  \\ 
            & $(1.50,\,2.00)$ & ~ $3.31 ~\pm~ 0.23$ ~  \\ 
            & $(2.00,\,3.01)$ & ~ $2.45 ~\pm~ 0.22$ ~  \\ 
            && ~ \\
            && ~ \\ \hline 
\end{tabular}
\vspace{0.15cm}
\caption{\it Values of the partial decay rates $\left[ \Delta \Gamma (q^2) \right]^{\rm EXP}$ for the $D \to \pi$ transition in the $q^2$-bins measured by  BESIII for the $D^0 \to \pi^- \ell \nu$~\cite{Ablikim:2015ixa} and $D^+ \to \pi^0 \ell \nu$~\cite{Ablikim:2017lks} channels.}
\label{tab:exp_data_gammaDpi_BESIII}
}
\quad
\parbox{0.475\linewidth}{
\centering
\begin{tabular}{|ccc|}
\hline
Experiment & $q^2 ~ (\mbox{GeV}^2)$ & $\left[ \Delta \Gamma(q^2) \right]^{\rm EXP} \cdot 10^{15} ~ (\mbox{GeV})$ \\ \hline \hline
 BESIII  $D^0$ & $(0.00,\,0.10)$  & ~ $5.807  ~\pm~  0.123 $ ~ \\
             & $(0.10,\,0.20)$  & ~ $5.762  ~\pm~  0.107 $ ~ \\
             & $(0.20,\,0.30)$  & ~ $5.466  ~\pm~  0.105 $ ~ \\
             & $(0.30,\,0.40)$  & ~ $4.987  ~\pm~  0.101 $ ~ \\
             & $(0.40,\,0.50)$  & ~ $4.933  ~\pm~  0.100 $ ~ \\
             & $(0.50,\,0.60)$  & ~ $4.248  ~\pm~  0.091 $ ~ \\
             & $(0.60,\,0.70)$  & ~ $4.086  ~\pm~  0.089 $ ~ \\
             & $(0.70,\,0.80)$  & ~ $3.637  ~\pm~  0.083 $ ~ \\
             & $(0.80,\,0.90)$  & ~ $3.313  ~\pm~  0.078 $ ~ \\
             & $(0.90,\,1.00)$  & ~ $2.982  ~\pm~  0.073 $ ~ \\
             & $(1.00,\,1.10)$  & ~ $2.618  ~\pm~  0.068 $ ~ \\
             & $(1.10,\,1.20)$  & ~ $2.192  ~\pm~  0.061 $ ~ \\
             & $(1.20,\,1.30)$  & ~ $1.864  ~\pm~  0.056 $ ~ \\
             & $(1.30,\,1.40)$  & ~ $1.508  ~\pm~  0.051 $ ~ \\
             & $(1.40,\,1.50)$  & ~ $1.145  ~\pm~  0.045 $ ~ \\
             & $(1.50,\,1.60)$  & ~ $0.866  ~\pm~  0.038 $ ~ \\
             & $(1.60,\,1.70)$  & ~ $0.565  ~\pm~  0.033 $ ~ \\
             & $(1.70,\,1.88)$  & ~ $0.250  ~\pm~  0.026 $ ~ \\ \hline             
BESIII $D^+$  & $(0.00,\,0.20)$  & ~ $11.18  ~\pm~ 0.40 $ ~ \\ 
             & $(0.20,\,0.40)$  & ~ $10.08  ~\pm~ 0.35 $ ~ \\ 
             & $(0.40,\,0.60)$  & ~~ $ 8.94  ~\pm~ 0.31 $ ~ \\ 
             & $(0.60,\,0.80)$  & ~~ $ 7.68  ~\pm~ 0.27 $ ~ \\ 
             & $(0.80,\,1.00)$  & ~~ $ 6.15  ~\pm~ 0.23 $ ~ \\ 
             & $(1.00,\,1.20)$  & ~~ $ 4.65  ~\pm~ 0.18 $ ~ \\ 
             & $(1.20,\,1.40)$  & ~~ $ 3.27  ~\pm~ 0.13 $ ~ \\ 
             & $(1.40,\,1.60)$  & ~~ $ 1.96  ~\pm~ 0.09 $ ~ \\              
             & $(1.60,\,1.88)$  & ~~ $ 0.67  ~\pm~ 0.05 $ ~ \\ \hline 
\end{tabular}
\vspace{0.15cm}
\caption{\it Values of the partial decay rates $\left[ \Delta \Gamma (q^2) \right]^{\rm EXP}$ for the $D \to K$ transition in the $q^2$-bins measured by BESIII  for the $D^0 \to K^- \ell \nu$ ~\cite{Ablikim:2015ixa} and $D^+ \to K^0 \ell \nu$~\cite{Ablikim:2017lks} channels.}
\label{tab:exp_data_gammaDK_BESIII}
}
\end{table}

In the case of the BELLE experiment~\cite{Widhalm:2006wz} the only available data are the values of $|V_{cx}| f_+^{D P}(q_i^2)$, which result from a variety of model-dependent shapes adopted for describing the vector form factor. 
The values of $|V_{cx}| f_+^{D P}(q_i^2)$ from the BELLE experiment are listed in Tables~\ref{tab:belle_data_fpVcd} and~\ref{tab:belle_data_fpVcs} for $D \to \pi$ and $D \to K$ semileptonic decays, respectively.

\begin{table}[htb!]
\parbox{0.4\linewidth}{
\centering
\begin{tabular}{|ccc|}
\hline
Experiment & $q_i^2 ~ (\mbox{GeV}^2)$ & $|V_{cd}| f_+^{D \pi}(q_i^2)$ \\ \hline 
 BELLE & $0.15$ & ~ $0.145  ~\pm~ 0.012$  ~ \\ 
             & $0.45$ & ~ $0.181  ~\pm~ 0.015$  ~ \\ 
             & $0.75$ & ~ $0.194  ~\pm~ 0.017$  ~ \\ 
             & $1.05$ & ~ $0.188  ~\pm~ 0.020$  ~ \\ 
             & $1.35$ & ~ $0.219  ~\pm~ 0.024$  ~ \\ 
             & $1.65$ & ~ $0.213  ~\pm~ 0.033$  ~ \\ 
             & $1.95$ & ~ $0.325  ~\pm~ 0.043$  ~ \\ 
             & $2.25$ & ~ $0.400  ~\pm~ 0.062$  ~ \\ 
             & $2.55$ & ~ $0.413  ~\pm~ 0.101$  ~ \\ 
             & $2.85$ & ~ $0.490  ~\pm~ 0.282$  ~ \\ 
            && ~ \\
            && ~ \\
            && ~ \\
            && ~ \\
            && ~ \\
            && ~ \\
            && ~ \\
            && ~ \\
            && ~ \\
            && ~ \\
            && ~ \\
            && ~ \\
            && ~ \\
            && ~ \\
            && ~ \\
            && ~ \\
            && ~ \\ \hline 
\end{tabular}
\caption{\it Values of $|V_{cd}| f^{D \pi}_+(q_i^2)$ from the BELLE collaboration~\cite{Widhalm:2006wz}, collected in Ref.~\cite{Rong:2014hea}.}
\label{tab:belle_data_fpVcd}
}
\qquad
\parbox{0.4\linewidth}{
\centering
\begin{tabular}{|ccc|}
\hline
Experiment & $q_i^2 ~ (\mbox{GeV}^2)$ & $|V_{cd}| f_+^{D K}(q_i^2)$ \\ \hline 
 BELLE & $0.10$ & ~ $0.688  ~\pm~ 0.029$  ~ \\ 
             & $0.17$ & ~ $0.762  ~\pm~ 0.029$  ~ \\ 
             & $0.23$ & ~ $0.743  ~\pm~ 0.029$  ~ \\ 
             & $0.30$ & ~ $0.811  ~\pm~ 0.032$  ~  \\ 
             & $0.37$ & ~ $0.762  ~\pm~ 0.032$  ~ \\ 
             & $0.43$ & ~ $0.817  ~\pm~ 0.036$  ~ \\ 
             & $0.50$ & ~ $0.856  ~\pm~ 0.039$  ~ \\ 
             & $0.57$ & ~ $0.915  ~\pm~ 0.039$  ~ \\ 
             & $0.63$ & ~ $0.882  ~\pm~ 0.039$  ~ \\ 
             & $0.70$ & ~ $0.798  ~\pm~ 0.039$  ~ \\ 
             & $0.77$ & ~ $0.996  ~\pm~ 0.042$  ~ \\ 
             & $0.83$ & ~ $0.970  ~\pm~ 0.045$  ~ \\ 
             & $0.90$ & ~ $0.921  ~\pm~ 0.045$  ~ \\ 
             & $0.97$ & ~ $1.015  ~\pm~ 0.052$  ~ \\ 
             & $1.03$ & ~ $1.070  ~\pm~ 0.052$  ~ \\ 
             & $1.10$ & ~ $0.911  ~\pm~ 0.055$  ~  \\ 
             & $1.17$ & ~ $1.083  ~\pm~ 0.065$  ~ \\ 
             & $1.23$ & ~ $1.067  ~\pm~ 0.068$  ~ \\ 
             & $1.30$ & ~ $1.219  ~\pm~ 0.078$  ~ \\              
             & $1.37$ & ~ $1.343  ~\pm~ 0.084$  ~ \\ 
             & $1.43$ & ~ $1.278  ~\pm~ 0.101$  ~ \\ 
             & $1.50$ & ~ $1.158  ~\pm~ 0.107$  ~ \\ 
             & $1.57$ & ~ $1.378  ~\pm~ 0.120$  ~ \\ 
             & $1.63$ & ~ $1.433  ~\pm~ 0.169$  ~ \\ 
             & $1.70$ & ~ $1.375  ~\pm~ 0.214$  ~ \\ 
             & $1.77$ & ~ $1.116  ~\pm~ 0.331$  ~  \\ 
             & $1.83$ & ~ $1.411  ~\pm~ 0.892$  ~  \\ \hline 
\end{tabular}
\caption{\it Values of $|V_{cs}| f^{D K}_+(q_i^2)$ from the BELLE collaboration~\cite{Widhalm:2006wz}, collected in Ref.~\cite{Fang:2014sqa}.}
\label{tab:belle_data_fpVcs}
}
\end{table}  
Though the Belle data on $|V_{cx}| f_+^{D P}(q_i^2)$ may contain some model dependence, they are considered for the extraction of $|V_{cx}|$ adopting the lattice determination of $f_+^{D P}(q^2)$ from Ref.~\cite{Lubicz:2017syv} at the center of each $q^2$-bin\footnote{We have checked that the exclusion of the Belle data from the analysis does not change significantly (within the errors) the extracted values of both $|V_{cd}|$ and $|V_{cs}|$.}.

\begin{figure}[htb!]
\includegraphics[scale=0.35]{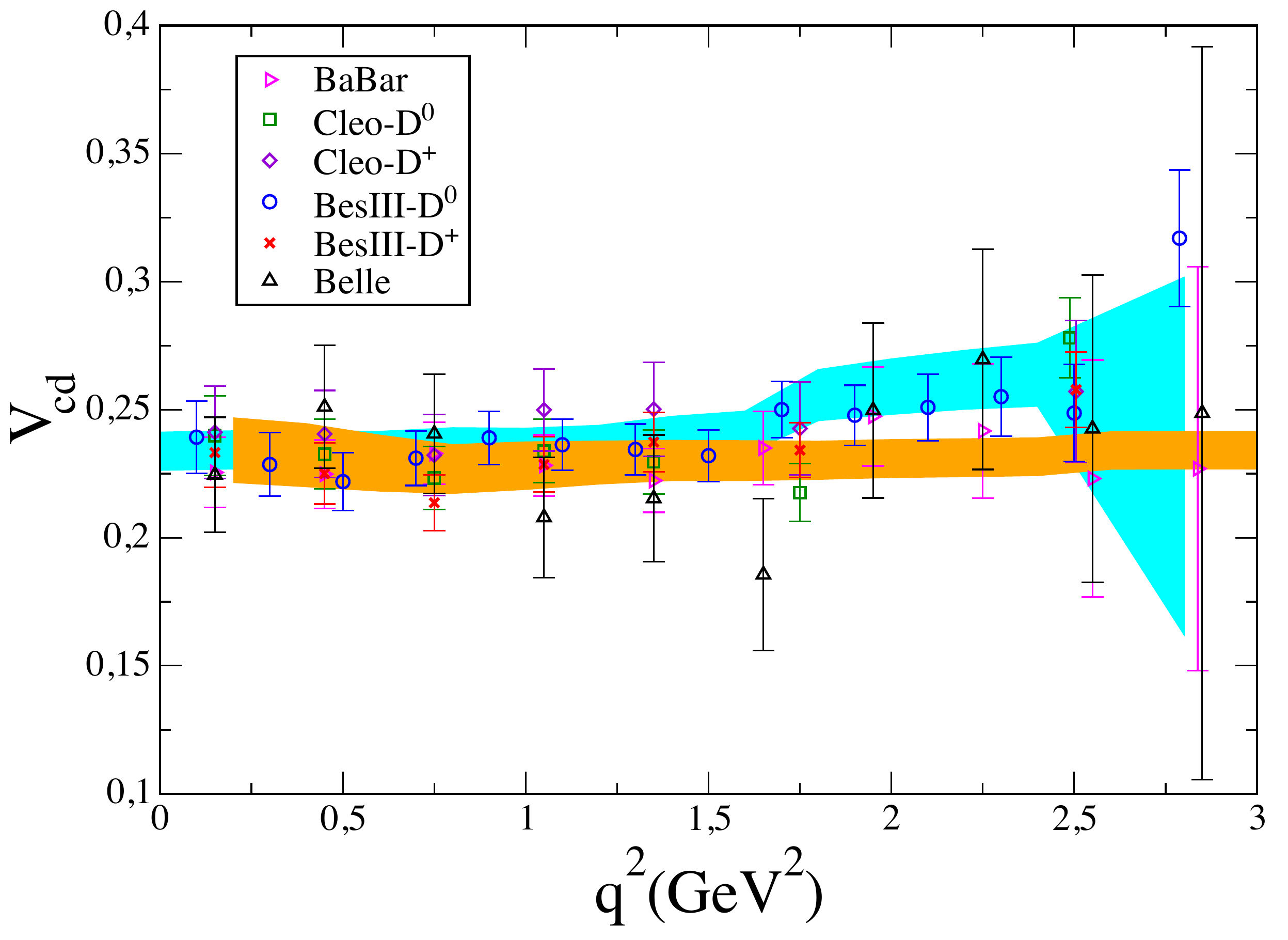} 
\includegraphics[scale=0.35]{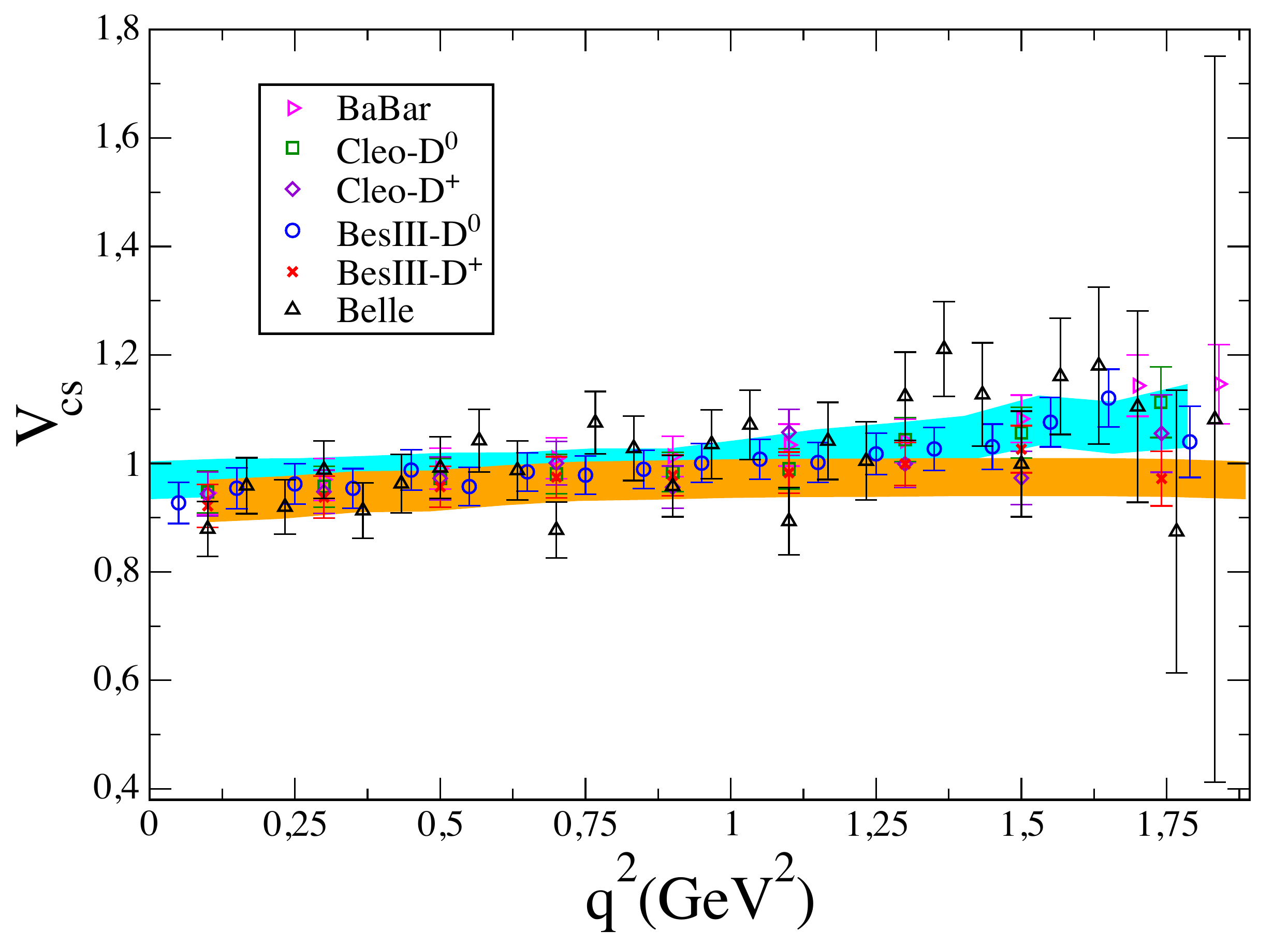}
\caption{\it Values of the CKM matrix elements $|V_{cd}|$ (left panel) and $|V_{cs}|$ (right panel) obtained by combining the theoretical predictions for the phase-space integrals $\left[ I(q_i^2) \right]^{\rm LAT}$, based on the vector form factor $f_+^{D P}(q^2)$ evaluated on the lattice in Ref.~\cite{Lubicz:2017syv}, with the experimental data of the $D \to \pi$ and $D \to K$ semileptonic differential decay rates $\left[ \Delta \Gamma(q_i^2) \right]^{\rm EXP}$, measured by BELLE \cite{Widhalm:2006wz}, BABAR \cite{Lees:2014ihu,Aubert:2007wg}, CLEO \cite{Besson:2009uv} and BESIII \cite{Ablikim:2015ixa,Ablikim:2017lks} collaborations. The bands are described in the text.}
\label{fig:Vcx_vs_q2}
\end{figure}

Combining the BABAR, CLEO and BESIII data from Tables~\ref{tab:exp_data_gammaDpi_BABAR}-\ref{tab:exp_data_gammaDK_BESIII} with the theoretical results for $\left[ I(q_i^2) \right]^{\rm{LAT}}$, and the BELLE data from Tables~\ref{tab:belle_data_fpVcd}-\ref{tab:belle_data_fpVcs} with the theoretical values for $f_+^{D P}(q_i^2)$, we get a determination of the CKM matrix element $|V_{cx}|$ for each experimental bin. 
The results are shown in Fig.~\ref{fig:Vcx_vs_q2} for both $|V_{cd}|$ and $|V_{cs}|$.
They exhibit an approximate constant behavior, except for $|V_{cs}|$ in the high-$q^2$ region, where some deviations are visible. 

The data are strongly correlated, since among the various $q^2$-bins the theoretical values of $\left[ I(q_i^2) \right]^{\rm LAT}$ as well as the values of $\left[ \Delta \Gamma (q_i^2) \right]^{\rm EXP}$ coming from the same experiment are correlated.
In order to take into account the correlations of the experimental data we have calculated the individual statistical+systematic covariance matrices for each experiments.  
Then, a global covariance matrix is obtained by combining in a block-diagonal form the separate covariance matrices given by BABAR, CLEO and BESIII collaborations in Refs.~\cite{Lees:2014ihu,Aubert:2007wg,Besson:2009uv,Ablikim:2015ixa,Ablikim:2017lks}. 
No covariance matrix is provided for the BELLE data on $|V_{cx}| f_+(q_i^2)$ in Ref.~\cite{Widhalm:2006wz}, which are treated as uncorrelated.
According to each individual experimental covariance matrix we generate bootstrap events for the values of the experimental partial decay rates $\left[ \Delta \Gamma (q_i^2) \right]^{\rm EXP}$.
Then, the bootstrap samplings of Ref.~\cite{Lubicz:2017syv} are used to evaluate the lattice values of $\left[I(q_i^2) \right]^{\rm LAT}$.
In this way the bootstrap distribution for $\left| V_{cx}(q_i^2) \right|^2$ is obtained and used to evaluate the final covariance matrix (always block-diagonal with respect to different experiments).

The values of $|V_{cd}|$ and $|V_{cs}|$ are determined using a constant fit and adopting a correlated-$\chi^2$ minimization procedure.
The uncertainties on $|V_{cd}|$ and $|V_{cs}|$ are not calculated using the obtained values of $\chi^2$. 
Instead, we use the bootstrap error related to the distribution of the extracted values of $|V_{cd}|$ and $|V_{cs}|$.
This ensures that all correlations and uncertainties are properly taken into account.
The results of the constant fit, including all $q^2$-bins, are 
  \be
     |V_{cd}| = 0.2341 ~ (74) \qquad , \qquad   |V_{cs}| = 0.970 ~ (33) ~ ,
     \label{eq:resultsCKM_ff}
 \ee
where the errors include all the various sources of statistical and systematic uncertainties of both the lattice form factors and of the experimental data.
We have estimated that $\approx 80 \%$ of the errors quoted in Eq.~(\ref{eq:resultsCKM_ff}) comes from the theoretical uncertainties and correlations.

In order to check the stability of these results we have also performed a series of constant fits including only the data below each given value of $q^2$, i.e., the number of data included in the fitting procedure increases as $q^2$ increases starting from $q^2 \simeq 0$.
The orange bands in Fig.~\ref{fig:Vcx_vs_q2} illustrate the results of these fits for $|V_{cd}|$ and $|V_{cs}|$ as a function of $q^2$.
The widths of the bands represent the spread of $|V_{cd}|$ and $|V_{cs}|$ at the level of one standard deviation.
It can clearly be seen that: ~ i) the variations of $|V_{cd}|$ and $|V_{cs}|$ are always within the uncertainties, and ~ ii) the uncertainties themselves do not change appreciably when the data at the highest values of $q^2$ are included.

We have also carried out a series of constant fits including only the data above each given value of $q^2$, i.e., the number of data included in the fitting procedure increases as $q^2$ decreases starting from $q^2 \simeq q_{max}^2$.
The cyan bands in Fig.~\ref{fig:Vcx_vs_q2} illustrate the corresponding results.
The extracted values of $|V_{cx}|$ differ from those corresponding to the orange bands only when the fits are restricted to the data at high values of $q^2$. 
The tension appears to be at the level of one standard deviation.
Once all the data are included in the fitting procedure (which for the cyan bands occurs close to $q^2 \simeq 0$ and for the orange bands close to $q^2 \simeq q_{max}^2$), the impact of the data at the highest values of $q^2$ is, however, negligible on the central values and errors given in Eq.~(\ref{eq:resultsCKM_ff}).

Our results (\ref{eq:resultsCKM_ff}) can be compared with those obtained in Ref.~\cite{Lubicz:2017syv} using the values of the vector form factor at $q^2 = 0$ and the experimental results for $|V_{cd} |f_+^{D \to \pi}(0)$ and $|V_{cs} |f_+^{D \to K}(0)$ provided by HFAG~\cite{Amhis:2016xyh}, namely
 \be
      |V_{cd}| = 0.2330 ~ (137) \qquad , \qquad |V_{cs}| = 0.945 ~ (38) .
      \label{eq:resultsCKM_q20}
 \ee
It turns out that the uncertainty of $|V_{cd}|$ in Eq.~(\ref{eq:resultsCKM_ff}) is smaller by $\approx 50\%$ with respect to the corresponding uncertainty in Eq.~(\ref{eq:resultsCKM_q20}), while for $|V_{cs}|$ the reduction of the uncertainty is marginal.
This is partially due to the higher degree of the correlations, found in Ref.~\cite{Lubicz:2017syv}, among the theoretical values of the vector form factor $f_+^{D P}(q^2)$ in the various $q^2$-bins in the case of the $D \to K$ transition with respect to the $D \to \pi$ one.
We stress that the same theoretical input from LQCD is used for describing the shape of the vector form factor $f_+^{D \pi(K)}(q^2)$ in all the experimental data, obtaining in this way a consistent SM analysis.
The impact of the above consistency might become more significant as the precision of LQCD calculations of the semileptonic form factors will be improved in the future.

Thus, the theoretical information on $f_+^{D P}(q^2)$ in the full $q^2$-range has allowed not only to guarantee a consistent extraction of $|V_{cd}|$ and $|V_{cs}|$ within the SM, but also to get a more precise determination of $|V_{cd}|$. 

\begin{figure}[htb!]
\includegraphics[scale=0.65]{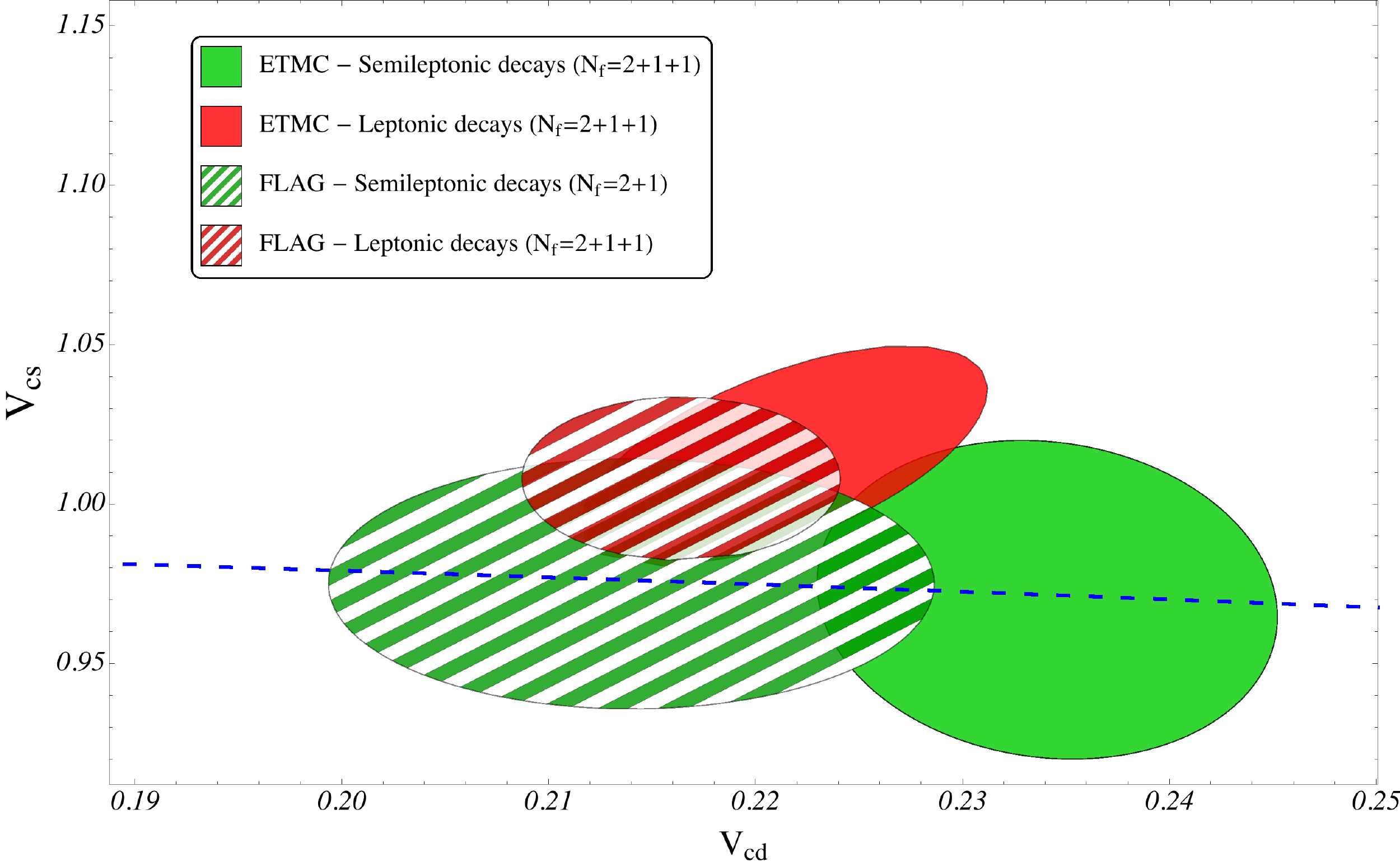}
\caption{\it Results for $|V_{cd}|$ and $|V_{cs}|$ obtained from leptonic $D$- and $D_s$-meson and semileptonic $D$-meson decays, represented respectively by red and green ellipses corresponding to a $68 \%$ probability contour. The solid ellipses are the results of Ref.~\cite{Carrasco:2014poa} and of this work, obtained with $N_f = 2 + 1 + 1$ dynamical quarks. The correlations between $|V_{cd}|$ and $|V_{cs}|$ are properly taken into account (see text). The striped ellipses correspond to the latest FLAG results~\cite{Aoki:2016frl}, which for the semileptonic decays are based on the LQCD results obtained in Refs.~\cite{Na:2011mc,Na:2010uf} with $N_f = 2 + 1$ dynamical quarks. The dashed line indicates the correlation between $|V_{cd}|$ and $|V_{cs}|$ that follows from the CKM unitarity.}
\label{fig:VcdVcs}
\end{figure}

Our semileptonic results (\ref{eq:resultsCKM_ff}) can be compared with the determinations $|V_{cd}| = 0.2221 (68)$ and $|V_{cs}| = 1.014 (25)$, obtained from the experimental $D$ and $D_s$ leptonic decay rates~\cite{Rosner:2015wva} adopting the ETMC results~\cite{Carrasco:2014poa} for the decay constants $f_D$ and $f_{D_s}$.
In Fig.~\ref{fig:VcdVcs} the above results from leptonic and semileptonic decays are reported as ellipses in the $(|V_{cd}|, ~ |V_{cs}|)$ plane corresponding to a $68\%$ probability contour. 
The ellipses corresponding also to the leptonic and semileptonic FLAG averages~\cite{Aoki:2016frl} for $|V_{cd}|$ and $|V_{cs}|$ are shown as well as the constraint imposed by the second-row unitarity is indicated by a dotted line. 
Note that in the case of both the leptonic and semileptonic FLAG results the correlation between $|V_{cd}|$ and $|V_{cs}|$ is not available, while in the case of the ETMC results the correlation between $|V_{cd}|$ and $|V_{cs}|$ have been properly considered, namely $\rho(|V_{cd}|, |V_{cs}|) \simeq 0.6$ in the case of leptonic decays and $\simeq -0.1$ for semileptonic decays.
 
Using $|V_{cb}| = 0.0360(9)$ from Ref.~\cite{Olive:2016xmw} and our semileptonic results (\ref{eq:resultsCKM_ff}) we can test the unitarity of the second row of the CKM matrix, obtaining
 \be
        \label{eq:utest}
        |V_{cd}|^2 + |V_{cs}|^2 + |V_{cb}|^2 = 0.996 ~ (64) ,
 \ee
which can be compared with the corresponding result $|V_{cd}|^2 + |V_{cs}|^2 + |V_{cb}|^2 = 0.949 (78)$ from Ref.~\cite{Lubicz:2017syv}.

Before closing this Section, we mention that very recently~\cite{Ablikim:2018frk} the BESIII collaboration has performed a test of the lepton universality (LU) in the $D \to \pi \ell \nu_\ell$ decays, obtaining
 \bea
       \label{eq:RLU_D0}
       {\cal{R}}_{LU}^{D^0\pi^-} \equiv \frac{{\cal{B}}(D^0 \to \pi^- \mu^+ \nu_\mu)}{{\cal{B}}(D^0 \to \pi^- e^+ \nu_e)} & = & 0.905 ~ (27)_{stat.} ~ (23)_{syst.} =
                                                          0.905 ~ (35) ~ , \\[2mm]
       \label{eq:RLU_D+}
       {\cal{R}}_{LU}^{D^+\pi^0} \equiv  \frac{{\cal{B}}(D^+ \to \pi^0 \mu^+ \nu_\mu)}{{\cal{B}}(D^+ \to \pi^0 e^+ \nu_e)} & = & 0.942 ~ (37)_{stat.} ~ (27)_{syst.} =
                                                            0.942 ~ (46) ~ . 
 \eea

Using in Eq.~(\ref{eq:Gamma}) the ETMC results for the semileptonic vector and scalar form factors of Ref.~\cite{Lubicz:2017syv} for both the $D \to \pi \mu(e) \nu$ and $D \to K \mu(e) \nu$ decays, we can predict for the first time both ${\cal{R}}_{LU}^{D\pi}$ and ${\cal{R}}_{LU}^{DK}$ within the SM using hadronic inputs calculated from first principles. 
Thanks to the strong correlation between the numerator and the denominator we get quite precise values in the isospin-symmetric limit of QCD, namely
 \bea
      \label{eq:RLU_DPi_SM}
      {\cal{R}}_{LU}^{D\pi} & = & 0.985 ~ (2) ~ , \\[2mm]
      \label{eq:RLU_DK_SM}
      {\cal{R}}_{LU}^{DK} & = & 0.975 ~ (1) ~ .
 \eea 
 The measured values (\ref{eq:RLU_D0})-(\ref{eq:RLU_D+}) are consistent with our SM prediction (\ref{eq:RLU_DPi_SM}) within $2.3\sigma$ and $1.2\sigma$, respectively.

\section{Conclusions}
 
We have presented a determination of the CKM matrix elements $|V_{cd}|$ and $|V_{cs}|$ obtained by combining the momentum dependence of the semileptonic vector form factors $f_+^{D \to \pi}(q^2)$ and $f_+^{D \to K}(q^2)$, recently determined from lattice QCD simulations, with the differential rates measured for the semileptonic $D \to \pi \ell \nu$ and $D \to K \ell \nu$ decays. 

Our analysis is based on the results for the semileptonic form factors produced by ETMC with $N_f = 2 + 1 + 1$ flavors of dynamical quarks in the whole range of values of the squared 4-momentum transfer accessible in the experiments~\cite{Lubicz:2017syv}. 
The statistical and systematic correlations between the lattice data as well as those present in the experimental data are properly taken into account.

With respect to the standard procedure based on the use of only the vector form factor at zero 4-momentum transfer, we obtain more precise and consistent results: $|V_{cd} |= 0.2341 ~ (74)$ and $|V_{cs} |= 0.970 ~ (33)$.
The second-row CKM unitarity is fulfilled within the current uncertainties:  $|V_{cd}|^2 + |V_{cs}|^2 + |V_{cb}|^2 = 0.996 ~ (64)$.

The results presented in this work depend crucially on the uncertainties and correlations of the input quantities, namely the experimental data available for the differential decay rates and the theoretical calculations of the semileptonic form factors of Ref.~\cite{Lubicz:2017syv}.
Future improvements of the precision of the above quantities are mandatory in order to assess the significance of any possible discrepancies with the Standard Model predictions, like in the case of the slight tension observed at high values of $q^2$ (see Fig.~\ref{fig:Vcx_vs_q2}).

Moreover, using for the first time hadronic inputs determined from first principles, we have calculated the ratio of the semileptonic $D \to \pi(K)$ decay rates into muons and electrons, which represent a test of lepton universality within the SM, obtaining in the isospin-symmetric limit of QCD: ${\cal{R}}_{LU}^{D\pi} = 0.985~(2)$ and ${\cal{R}}_{LU}^{DK} = 0.975~(1)$.

\section*{Acknowledgements}
We warmly thank our colleagues V.~Lubicz and C.~Tarantino for useful discussions and a careful reading of the manuscript.

\bibliographystyle{JHEP}

\bibliography{rifbiblio}

\end{document}